# Safety-Critical Adaptation in Self-Adaptive Systems

Simon Diemert
Department of Computer Science
University of Victoria
Victoria, Canada
sdiemert@uvic.ca

0000-0001-9493-7969

Jens Weber
Department of Computer Science
University of Victoria
Victoria, Canada
jens@uvic.ca

*Abstract* — Modern systems are designed to operate in increasingly variable and uncertain environments. Not only are these environments complex, in the sense that they contain a tremendous number of variables, but they also change over time. Systems must be able to adjust their behaviour at run-time to manage these uncertainties. These "self-adaptive systems" have been studied extensively. This paper proposes a definition of a *safety-critical self-adaptive system* and then describes a taxonomy for classifying adaptations into different types based on their impact on the system's safety and the system's safety case. The taxonomy expresses criteria for classification and then describes specific criteria that the safety case for a self-adaptive system must satisfy, depending on the type of adaptations performed. Each type in the taxonomy is illustrated using the example of a safety-critical self-adaptive water heating system.

I. INTRODUCTION

Consider an autonomous parcel delivery robot that operates in an urban environment. The robot navigates pedestrian sidewalks to deliver a parcel to the intended address. To complete its mission, the robot must detect sidewalk boundaries and determine its position on the sidewalk, a task often referred to as "localization." To be economically viable, the robot must be deployable in a wide range of urban centers, each with unique sidewalk characteristics. Of course, because they are built by humans, sidewalks are highly variable structures with inconsistent boundaries. Thus, the robot's sidewalk localization function must handle uncertainty arising a variable environment.

A common strategy for managing uncertainty is to constrain the operating environment (also referred to as the "operational domain") of the system. The system is only permitted to operate within environments that satisfy a set of pre-determined constraints. This is typically enforced by a combination of human supervision and run-time checking of the environment by the system. Constraining the operational domain reduces the amount of uncertainty arising from environmental variability. For example, the parcel delivery robot might be restricted to operating on fair weather days on a selection of streets in the downtown core of a specific city, where the sidewalks have a consistent size and appearance. Restricting the operational domain can severely limit the usefulness of the system. However, even within a restricted operating environment a significant amount of uncertainty might exist.

To complicate matters, operating environments are rarely static: they change over time! For the robot, the sidewalk materials wear out, new materials are used to make new sidewalks or repair damage, and construction regulations change. For example, suppose a city decides to paint the edges of some sidewalks bright red to indicate "no stopping" zones for automobiles. Such a change could occur quickly with little warning. As a result, the robot might not be able to detect the sidewalk boundary.

To mitigate uncertainty arising from variable and changing environments, the robot should be able *adapt* to its environment. For example, the component responsible for localizing the robot on a sidewalk might have several configurations or profiles that are tuned for different sidewalk types and weather conditions. At run-time, the robot automatically selects the appropriate configuration for the current conditions. The robot might also contain functions responsible for learning about its environment and change its configuration as the environment changes over time. Performing these adaptations would improve the availability of the robot since it could be deployed in a much wider range of operating environments and is less likely to encounter problems it cannot handle at run-time. Systems that adjust their behaviour at run-time to adapt to variable or changing operating conditions are called *self-adaptive systems* [1]. Self-adaptation is an important strategy for managing run-time uncertainty.

The parcel delivery robot, and its sidewalk localization function, are also an example of a *safety-critical system*. That is, misbehaviour of the system might cause harm to persons, property, or the environment [2]. For example, a failure to detect the boundary of a sidewalk on a busy street might cause the robot to drive into traffic and contribute to an accident. Safety is a unique system property that is often in tension with other properties such as availability or performance (the safest parcel delivery robot is one that never leaves the garage) [2].

Assuring self-adaptive systems and establishing their "trustworthiness" continues to be an open challenge [1, 3, 4, 5]. Thus far, research in the area of "assurances" for self-adaptive systems has focused on assuring properties of self-adaptive systems in general terms in the sense that the methods can be applied to any property, including safety, reliability, availability, or performance. Many authors have described methods for assuring self-adaptive systems that can be applied to safety, notably the ENTRUST framework [6]. However, to our knowledge, no work addresses safety at a conceptual level and answers the question: *what does it mean for a self-adaptive system to be safe?* To fully realize the potential of self-adaptation in safety-critical applications such as transportation, energy generation, or medical systems, additional work is required to develop a definition of safety, independently of other system properties, for self-adaptive systems.

This concept paper formalizes the notion of safety-critical self-adaptive systems by making two contributions. First, definitions for the terms *safety-critical self-adaptive system* and *safe adaptation* are proposed. Second, the definition is used by a taxonomy for classifying adaptations performed by a self-adaptive system, from a safety perspective. Central to the taxonomy is the relationship between the adaptations performed by the system and a safety case that describes the rationale for why the system is believed to be safe. The taxonomy consists of four "types" that successively describe an increasingly dynamic relationship between adaptations and the system's safety case. Each type is illustrated using the example of a simple water heater. To the authors' knowledge, this paper contains the first formalized description and taxonomy of safety-critical self-adaptive systems.

The remainder of this paper is structured as follows. The Background section provides foundational details on self-adaptive systems, safety-critical systems, and safety cases. Next, the main contribution of the paper, the definitions of safety-critical self-adaptive system and safe adaptation are provided, and the taxonomy for classifying adaptations is described. Then, related work in the area of self-adaptive systems is reviewed. Finally, the paper closes with a short description of next steps for this research.

## II. Background

This section provides foundational definitions and details about self-adaptive systems, safety-critical systems, and safety cases.

### A. Self-Adaptive Systems

As described above, a self-adaptive system is one that adjusts its behaviour in response to changes in the system's operating conditions [1]. Weyns provides a two-part definition of a self-adaptive system that captures these ideas, the definition is expressed as two principles that a system must satisfy in order to be considered self-adaptive:

> *"External Principle – A self-adaptive system is a system that can handle changes and uncertainties in its environment, the system itself, and its goals autonomously (i.e., without or with minimal required human intervention)."*
>
> *Internal Principle – A self-adaptive system comprises two distinct parts: the first part interacts with the environment and is responsible for the domain concerns (…); the second part consists of a feedback loop that interacts with the first part (and monitors its environment) and is responsible for the adaptation concerns."* [1].

The relationship between self-adaptive systems and autonomous systems is worth further discussion. The notion of *autonomy* appears as part of the External Principle and so autonomy is a necessary condition for a self-adaptive system in so far as the system decides (on its own) when to adjust its behaviour. However, the level autonomy may be described on a scale: some systems are more autonomous than others and are able to handle more uncertainty with less human intervention. Weyns' definition does not describe the extent of autonomy required. Autonomy is regarded as a separate dimension from adaptation [4]. Therefore, it is possible for a system to exhibit a modest degree of autonomy (e.g., an adaptive thermostat) while still being self-adaptive. The reverse is not true: many autonomous systems are not self-adaptive. For example, if the parcel delivery robot is deployed with a single configuration and has no capacity to adjust its behaviour, then it would not be considered self-adaptive.

The Internal Principle requires that a self-adaptive system adopt a high-level architecture comprised of two major sub-systems. These are referred to as the *managed system* and *managing system* [1]; the relationship between these sub-systems is shown in Figure 1 and is discussed below.

The managed system performs the primary function of the system. In the case of the parcel delivery robot the managed system navigates to the intended recipient to deliver a package and includes performing tasks like sidewalk localization. The managed system must provide interfaces for the managing system to perform adaptations. For example, the parcel delivery robot's sidewalk localization function might listen for network requests telling it to switch to a different internal configuration. The internal architecture of the managed system is application specific and is not discussed further.

The managing system monitors the managed system and the environment and adjusts the configuration of the managed system as needed to satisfy one or more adaptation goals prescribed by the system's designers or operators. Adaptation goals describe non-functional requirements for the system and might relate to availability, reliability, performance, or efficiency. For example, the parcel delivery robot might have a performance oriented adaptation goal to complete two deliveries per operating hour.

A common strategy is to implement the managing system using a Monitor-Analyze-Plan-Execute and Knowledge (MAPE-K) reference model [1]. The Monitor component records observations of the system and the environment and maintains one or more models of the current state that are stored in a shared Knowledge Repository component. The Analyzer component observes the current state and determines whether the prescribed adaptation goals are satisfied; if they are not satisfied, then the Analyzer generates candidate "adaptation options" (new configurations for the managed system). The Planner component selects the best adaptation option and determines a plan for transitioning the system between the existing system configuration and new configuration. Finally, the Executor applies the selected adaptation option to the managing system in accordance with the Planner's chosen plan. See Chapter 4 in [1] for a detailed description of the MAPE-K reference architecture.

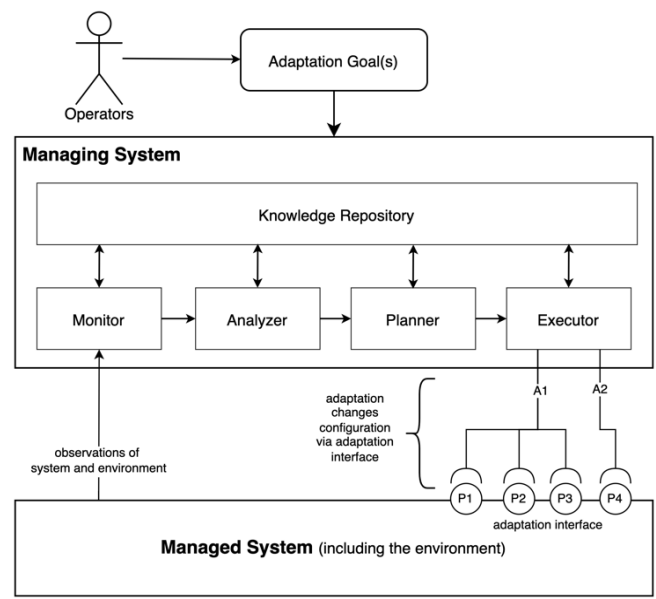

*Figure 1 - Architecture of a self-adaptive system.*

An *adaptation option* defines the system configuration "that can be reached from the current configuration by adapting the managed system" [1]. The managing system analyzes candidate adaptation options and selects the best option (per adaptation goals) to apply to the managed system. An *adaptation action* is a "unit of activity that needs to be executed on the managed system in order to adapt it" [1]. An adaptation action defines a sequence of instructions performed on the managed system to change its configuration.

The taxonomy described below in Section IV aims to classify adaptations performed by the managing system. However, the term "adaptation" is somewhat ambiguous and does not appear to be precisely defined in the self-adaptive systems literature. Consider the sentence: *The adaptation performed by the managing system changes parameter X of the managed system*. In this sentence it is not clear if the word "adaptation" describes a specific instance of change performed on the managed system (e.g., executing an adaptation action to set X to a specific value identified in an adaptation option) or if it refers to a generic description (i.e., a template) for a change that could be executed on the managed system (i.e., a capability of the managed system). Though this ambiguity provides flexibility for authors writing about self-adaptive systems, it also leads to difficulties when attempting to classify adaptations.

In this paper we favour the second interpretation described above. More precisely, we use the term *adaptation model* to mean a generic specification for the change that the managing system executes (or is capable of executing) on the managed system. An adaptation model identifies the parameters in the managed system's configuration that may be changed and defines the set of constraints for the parameters. An adaptation option describes a specific realization of an adaptation model. It is possible for a self-adaptive system to have multiple adaptation models, each describing a different type of change. In Figure 1 two adaptation models are sketched (A1 and A2): model A1 identifies parameters P1, P2, and P3 and model A2 identifies only parameter P4. Though not depicted, the adaptation models could define constraints for the parameters. For example, if P1 and P2 are numerical parameters, then A1 might require $P1 > a$ if $P2 > b$ for values of $a$ and $b$. In the remainder of this paper, unless otherwise indicated, the term "adaptation" should be regarded as synonymous with an "adaptation model".

*B. Safety-Critical Systems*

There are a multitude of definitions for terms related to system safety, although they all generally describe the same ideas. This paper adopts the definitions due to Leveson. An *accident* is an "undesired and unplanned event that results in a loss (including loss of human life or injury, property damage, environmental pollution, and so on)" [2] and *safety* is defined as "freedom from accidents (loss events)" [2]. Since the definition of safety depends on the occurrence of an accident, it follows that safety is a property that must be assessed at the whole system-level. Unlike other properties, such as reliability or performance, which can be reasoned about at the component level, safety must be discussed in the context of the environment that the system interacts with. That is, safety must be considered in the context of an environment in which accidents occur.

Safety engineers define *hazards* to bridge between the world of system operations/functions and the real-world environment. A *hazard* is defined as "a system state or set of conditions that, together with a particular set of worst-case environment conditions, will lead to an accident (loss)" [2]. The identification, assessment, and mitigation of hazards is an essential step in internationally recognized safety standards such as ISO 26262 (automotive electronics and software) and IEC 61508 (industrial control systems) [7, 8].

Combining these definitions, a system (or function) is *safety-critical* if its incorrect, inadvertent, out-of-sequence, or inadvertent action in combination with conditions in the environment might contribute to the occurrence of, or fail to mitigate the effect of, a hazard occurrence or accident. This definition was adapted from [9, 10].

*C. Safety Assurance Cases*

An *assurance case*, is defined as a "reasoned and compelling argument, supported by a body of evidence that a system, service, or organization will operate as intended for a defined application in a defined environment" [11]. Assurance cases may argue about any property of interest (reliability, security, performance, etc.). This paper is focused on safety so the case is referred to as a *safety assurance case* or simply *safety case*. The objective of a safety case is to argue that the system is acceptably safe for its intended application.

Kelly developed the modern concept of a safety case in the late 1990s [12]. They have since become an important safety engineering method in several industries. For instance, in the automotive industry, safety cases are required for compliance with ISO 26262 – *Road vehicles – Functional safety* [7] and UL 4600 – *Standard for the Safety of Autonomous Products* prescribes the content of safety cases that argue about autonomous vehicle safety [13].

Safety cases are often expressed using a tree-like notation such as Goal Structuring Notation (GSN) [12, 11]. Nodes in the tree contain: goals - claims to be argued, strategies – describing the argument approach, or solutions – describing supporting evidence. Other complementary notations for expressing safety cases exist, such as Claims-Argument-Evidence notation [14] or Eliminative Argumentation [15, 16]. Though popular, structured notations are not strictly necessary: narrative ("essay-style") safety cases are also used to express safety arguments.

When a safety case is written down (using either a notation or narrative form) it is referred to as an "explicit" safety case. These are advantageous since they allow stakeholders to directly reason about safety. However, even if a safety case is not written down, the rationale for safety still exists in the minds of the stakeholders. This rationale is sometimes referred to as an "implicit safety argument" or "implicit safety case"[1] [17, 18]. For example, in industries where standards do not mandate the documentation of a safety case, a collection of artifacts (test results, analysis results, inspection reports, etc.) are generated by carrying out an established engineering process, but a system-specific rationale for why these artifacts are relevant is not clearly documented. The notion of an implicit safety case is important for this paper because the data (particularly run-time analysis results) stored within the Knowledge Repository of a self-adaptive system can, in some cases, be regarded as an implicit case justifying the decision by the system to apply an adaptation option.

---

[1] The term "implicit safety case" is self-contradicting. A safety case demands that an argument be documented. Regardless, it is useful for describing situations where the argument is widely known but not documented.

### III. SAFETY-CRITICAL SELF-ADAPTIVE SYSTEMS

Using the definitions introduced in Section II, this section defines a safety-critical self-adaptive system and then uses that definition to define a safe adaptation.

#### A. Definition of a Safety-Critical Self-Adaptive System

A *safety-critical self-adaptive system* is a system that meets all of the following criteria:

- The system satisfies Weyns' External Principle and can "handle changes and uncertainties in its environment, the system itself, and its goals autonomously" [1].

- The system satisfies Weyns' Internal Principle and is comprised of a managed system and managing system where the managed system performs the primary system function(s) and the managing system adapts the managed system.

- The managed system performs (or is capable of performing) functions that are safety-critical, i.e., its incorrect, inadvertent, out-of-sequence, or inadvertent action in combination with conditions in the environment might contribute to the occurrence of, or fail to mitigate the effect of, a hazard occurrence or accident.

This definition is simple in the sense that it combines the definitions of self-adaptive and safety-critical systems. However, it leads to additional observations about the nature of these systems.

First, the criterion for safety-criticality is assessed only for the managed system. This follows because it is the managed system that is strictly responsible for performing the system function and interacting with the environment. Though the managing system might contribute to a hazard occurrence (e.g., by selecting an inappropriate adaptation option), it is the managed system that performs (or fails to perform) an action leading to a hazard. A consequence of this is observation is that designers performing hazard identification, an essential safety engineering activity, may focus their attention on the interface between the managed system and the environment.

Second, the criteria do not constrain the nature of adaptation(s) performed by the managing system. The adaptations might (or might not) impact the safety of the managed system without changing whether the system is regarded as safety-critical. This has implications for how adaptations are reasoned about: all adaptations must be analyzed to determine whether they are safe. This leads to the question: *what does it mean for an adaptation to be safe?*

#### B. Definition of a Safe Adaptation

The definition of a safe adaptation combines two ideas. First, as noted above, in a safety-critical self-adaptive system the managed system is the source of hazard occurrences because it is responsible for interfacing with the environment. This suggests that the definition of a safe adaptation should be focused on the managed system. Second, the managing system uses adaptation interfaces to change the configuration of the managed system according to the chosen adaptation option. This suggests that the definition of safe adaptation should consider the safety of the related adaptations options and the corresponding adaptation actions.

A *safe adaptation option* is an adaptation option that, when applied to the managed system, does not result in, or contribute to, the managed system reaching a hazardous state. A safe *adaptation action* is an adaptation action that, while being executed, does not result in or contribute to the occurrence of a hazard. It follows that a *safe adaptation* is one where all adaptation options and adaptation actions are safe.

These definitions of safety-critical system adaptative system and safe adaptation lead to a further question: *how do we determine if an adaptation is a safe adaptation?* This question is addressed by the taxonomy described in the next section.

### IV. TAXONOMY FOR SAFETY-CRITICAL ADAPTATION

As described above, a safety-critical self-adaptive system might be capable of performing multiple different adaptations (i.e., has multiple adaptation models). In some cases, adaptations will not (or are unlikely to) affect the safety of the overall system. For example, a voice control function in the parcel delivery robot might adapt to a local accent or language dialect. But adapting the language interface is unlikely to impact a safety-critical function like the robot's trajectory control algorithm(s). But in other cases, an adaptation might alter the system's behaviour of safety-critical functions. For example, some robot control systems might adapt to different weather conditions to improve performance properties, such as delivery time.

A safety case describes reasons that a system is believed to be safe. Historically, safety cases have been regarded as static artifacts, developed prior to deployment. However, safety cases have been recognized as an important part of assurance for self-adaptive systems [3]. Several authors have proposed "dynamic" safety cases that are updated after deployment to reflect the latest data gathered from the field [5, 19, 20]. In some self-adaptive systems, the safety case is regarded as part of the Knowledge Repository maintained by the managing system. Since a self-adaptive system adapts its behaviour at run-time, then (depending on the nature of the adaptation) the safety case for the system can also be subject to adaptation. The effect of the adaptation on the safety case, in combination with the effect of the adaptation the managed system, can be used as a basis for classifying adaptations.

This section presents a taxonomy for classifying adaptations according to their impact on the system's safety-critical function(s) and on the safety case. Four "types" of adaptations are identified: Type 0, Type I, Type II, and Type III. For each type criteria are described that may be used to classify an adaptation. In the narrative below, criteria are denoted by curly braces and an identifier of the form {Tx.Cy} where *x* identifies the Type (0, I, II, III) and *y* identifies the criteria number. All criteria for a given type must be satisfied for an adaptation to be classified as such.

Regardless of whether an adaptation is thought to affect safety-critical functions, all adaptations that the managing system is capable of performing must be classified. Once an adaptation is classified, the taxonomy defines obligations that the safety case for the system must satisfy in relation to that adaptation. Similarly to the criteria, obligations are denoted by {Tx.By}. The set of criteria and obligations are summarized in **Error! Reference source not found.** at the end of this section for convenience.

A running example of a self-adaptive water heating system is used to illustrate each classification type. The system is depicted in Figure 2 and is used to heat water to a setpoint provided by an operator. The managed system is supplied with water with a variable temperature and inflow rate. A proportional-integral-derivative (PID) feedback controller controls a heating element that is used to heat the water to the setpoint. The outflow rate of the water is equal to the inflow rate so that the heater's reservoir does not overflow. For this hypothetical system, a safety hazard occurs when the temperature of the outflowing water exceeds 90 °C for more than 2 seconds.

The managing system observes the behaviour of the managed system and adapts PID controller configuration (proportional, integral, and derivative gains) to achieve an adaptation goal: when the setpoint is increased, the outflow temperature should reach that setpoint is less than 60 seconds. Changing the configuration of the controller constitutes an adaptation. The managing system uses a combination of conventional analysis and learning methods to determine when adaptation is necessary and to select adaptation options (controller configurations).

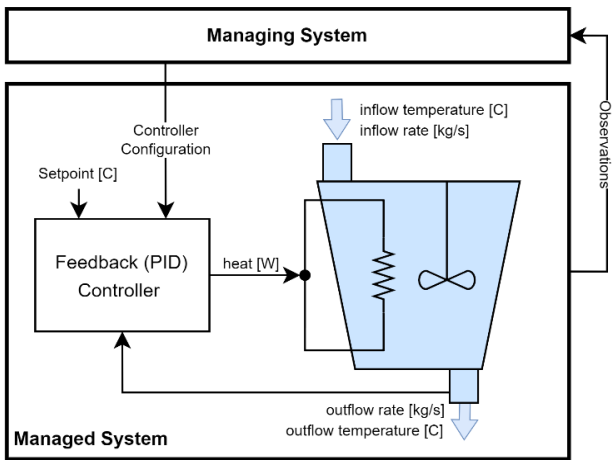

*Figure 2 - Running example: self-adaptive water heater.*

### A. Type 0 Adaptations – Non-Interference

Adaptations that do not affect the safety-critical functions of the managed system are classified as Type 0{T0.C1}. The safety case for the system must argue that a Type 0 adaptation does not interfere with the safety-critical functions {T0.B1}. Since the adaptation is separate from the safety-critical functions, the safety case for a Type 0 adaptation is static and, after it is developed at design-time, does not change {T0.B2}.

A "safety monitor" (sometimes called a "safety bag" or "safety supervisor") is a widely used system architecture for mitigating hazard occurrences in conventional safety-critical systems. This architecture may be used when the main system function implemented by the managed system can be separated from the task of detecting and responding to hazardous conditions. If a safety monitor is used and is argued in the safety case to be independent of the adaptations, then the adaptations on the other (non-safety monitor) elements of the system can be classified as Type 0 adaptations.

*1) Type 0 Example:* In the water heater example, an independent safety monitor might measure the temperature of the outflowing water. If the outflow temperature exceeds 90 °C, then the safety monitor closes a valve and disables power to the main water heater within 2 seconds, thus avoiding a hazard occurrence. This modified system is depicted in Figure 3. The safety case for this system must argue that: 1) the adaptations performed by the managing system do not affect the ability of the safety monitor to respond to hazardous temperature conditions, and 2) the safety monitor reliably measures and responds to hazardous outflow temperatures.

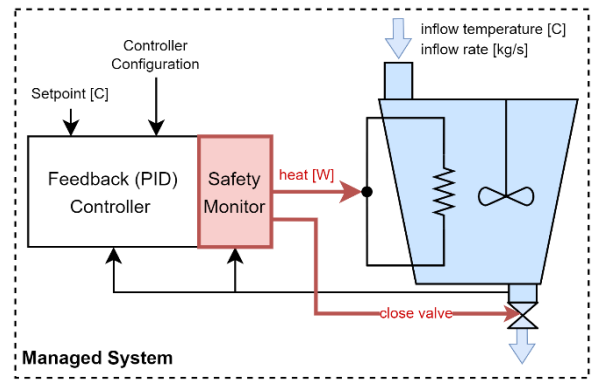

*Figure 3 - Illustration of safety monitor architecture for Type 0 adaptation (only managed system shown).*

Though it is widely used system architecture, the use of a safety monitor is not a necessary condition for Type 0 adaptations. Other approaches may be used to establish non-interference. For instance, the previously mentioned natural language interface for the parcel delivery robot could be argued as independent if the language processing system was realized on a wholly independent computer and network from the robot's motion control systems (i.e., an "air gap").

### B. Type I Adaptations – Static Assurance

Type I adaptations affect safety-critical functions of the managed system {TI.C1}. If an inappropriate adaptation option is selected or the adaptation action is executed incorrectly, then it is possible for a hazard to occur. However, an important property of a Type I adaptation is that all adaptation options are defined at design-time, before the system is deployed {TI.C2} and that these adaptation options are shown at design-time to be safe for the entire operational domain of the managed system {TI.C3}. That is, for any adaptation option applied, the attainable state(s) of the system are known (at design-time) to be safe.

The safety case for a self-adaptative system must argue three points related to a Type I adaptation. First, the safety case must argue that the managing system will only select and apply the adaptation options that were defined at design-time {TI.B1}. Second, the safety case must argue that each adaptation option is safe within the entirety of the managed system's operational domain {TI.B2}. Finally, the safety case must argue that execution of the adaptation action by the managing system is safe {TI.B3}.

Like a Type 0 adaptation, the safety case for a Type I adaptation is static and does not change at run-time {TI.B4}. This is in contrast to Type II and III adaptations that change the safety case at run-time. But, for a Type I adaptation, since all of the adaptation options are defined and shown to be safe at design-time, there is no need for the managing system to maintain or adapt a safety case at run-time.

*1) Type I Example:* To illustrate Type I adaptation, consider an instance of the water heater system shown in Figure 2 where the managing system has a pre-determined set of 10 PID controller configurations that form the set of adaptation options that may be executed on managed system. At design time, each of these adaptation options has been shown using a combination of analysis and simulation, to produce controller outputs that do not result in the outflow water temperature exceeding 90 °C for more than 2 seconds. The safety case for the system incorporates the analysis and simulation results as evidence that the adaptation options are safe. The safety case also contains arguments and evidence to support the claim that the managing system will execute adaptation options safely and will not use an adaptation option that is not one of the 10 pre-determined options.

In this example, the choice of 10 pre-determined adaptation options is arbitrary. Provided the options are defined and each option is shown at design-time to be safe, there is no reason that a Type I adaptation cannot have an infinite number of adaptation options. For instance, the choice of PID controller gains might be defined over an infinite subset of the Real numbers.

*C. Type II Adaptations – Constrained Assurance*

Type II adaptations occur in self-adaptive systems that seek to optimize their behaviour at the expense of flexibility in their operating environment. Like Type I adaptations, Type II adaptations affect safety-critical functions of the managed system {TII.C1} and the set of adaptation options are defined at design-time {TII.C2}. Furthermore, all adaptation options are shown at design-time to be safe, but the safety of each depends on additional constraints (or assumptions) imposed about the operational domain {TII.C3}. At run-time, part of the adaptation action executed by the managing system involves updating a dynamic safety case to include the constraints associated with the chosen adaptation option {TII.C4}. While the initial safety case for the system covers a broad operational domain, the adapted safety case is only valid within the constrained operational environment. Changing the operational environment such that the constraints are no longer satisfied might result in unsafe behaviour of the adapted system. Additionally, each Type II adaptation must monotonically increase the operational domain constraints {TII.C5}.

Similarly to Type I adaptation, the static portion of a safety case for a Type II adaptation must argue that: only the defined adaptation options are executed by the managing system {TII.B1}; the defined adaptation options are safe subject to assumptions on the operational domain {TII.B2}; and the managing system safely executes adaptation options {TII.B3}. These obligations describe static elements of the safety case that are included in the case at design-time and are not subject to any adaptation at run-time.

However, an essential property of Type II adaptation is that the safety case for the system is dynamic and is changed as part of the adaptation operation {TII.C4}. Therefore, the dynamic elements of the safety case must address two additional points. First, the safety case must argue that if an adaptation is applied, then the operational domain satisfies the imposed constraints {TII.B4}. Such an argument might depend on observations of the operational domain collected at run-time by the managing system. Second, the safety case must argue that violations of the current operational constraints (due to phenomena beyond the system's control) are either are safely handled or occur with an acceptably low probability {TII.B5}. Since the safety case is dynamic and adapted at run-time, it should be regarded as part of the "knowledge" maintained by the managing system. Ideally the safety case is explicitly described using a notation that can be operated upon by the managed system (e.g., using Goal Structuring Notation) and a human operator of the system should be able to query the managing system's knowledge base to determine the current operational constraints imposed on the system. However, for some systems, an implicit safety case consisting of data and analysis results collected by the managing system might be preferred.

*1) Type II Example:* To illustrate Type II adaptation, consider yet another instance of the water heater system. As before, this instance has a pre-determined set of 10 adaptation options. At design-time, each of these options has been demonstrated to be safe *subject to conditions on the inflow rate and inflow temperature*. The water heater is designed to be initially deployed with adaptation option #1 enabled, which has a very conservative controller configuration that has been shown to be safe for a wide range of water inflow rates and temperatures. The dynamic safety case initially reflects these permissive operating conditions. After deployment, the managing system observes of the inflow rate and temperature and makes adaptation decisions when there is high confidence that a set of operational constraints is satisfied.

Suppose that the water heater is deployed in an environment that has cold and fast moving inflow water (e.g., a northern community). One of the adaptation options, option #9, is a very "aggressive" controller configuration with large PID gains that is suitable for cold (less than 2 °C) water with a fast inflow rate. Before executing an adaptation, the managing system collects data on the inflow water for an extended period of time and performs a statistical analysis which confirms with high confidence that inflowing water is consistently fast flowing and below 2 °C. The managing system applies adaptation option #9, this includes: 1) changing the PID controller gains, 2) updating operational constraints in the dynamic portion of the safety case stored with the managing system's knowledge repository, and 3) updating the safety case with observations and analysis results that demonstrate that operational constraints are satisfied.

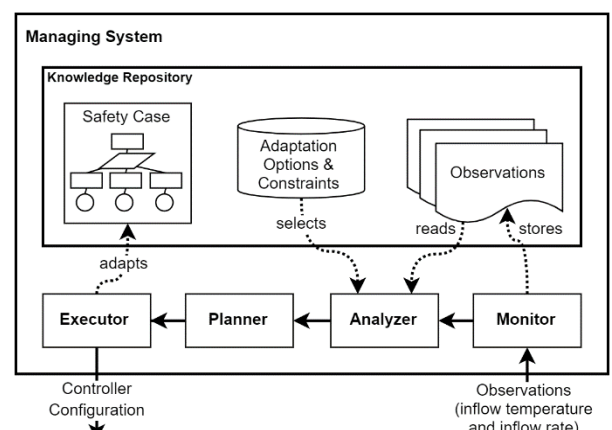

*Figure 4 – Managing system for Type II adaptation.*

*D. Type III Adaptations – Dynamic Assurance*

Like Type I and II adaptations, Type III adaptations affect the safety-critical functions of the managed system {TIII.C1}. However, unlike Type I and II systems, the set of adaptation options for a Type III adaptation is not defined at design-time {TIII.C2}. While it might be theoretically possible to define a set of adaptation options, it might not be reasonable to do so in practice. For instance, if the parameter space of possible adaptation options is very large, then it might not be feasible to define the set of adaptation options. A consequence of being unable to define the set of adaptation options is that adaptation options cannot be shown to be safe at design-time, prior to deployment. Instead, adaptation options are assessed at run-time by the managing system and the results are incorporated into a dynamic safety case as part of the adaptation operation performed by the managing system {TIII.C4}. The safety case is stored as part of the knowledge repository within the managing system {TIII.C5}. However, since assessment of adaptation options is performed at run-time, there will be uncertainty about whether the chosen adaptation option is actually safe; this uncertainty arises from limitations of automated assessment performed by the managing system {TIII.C6}.

The safety case for Type III adaptation should adopt the notion of "Positive Trust Balance" (PTB) proposed by Koopman [21]. This approach acknowledges that autonomous and adaptive systems are subject to many sources of uncertainty that cannot reasonably be addressed at design-time. After rigorous (but necessarily imperfect) design-time efforts are made, the system may be deployed and closely monitored using safety performance indicators (SPIs) [21, 22] The validity of the safety case should be regularly re-assessed based on data provided by the SPIs.

By adopting the PTB approach, obligations for a Type III safety case can be identified. The static (design-time) portion of the safety case must address five items. First, the safety case must argue that the managing system safely executes the adaptation action {TIII.B1}. Second, the safety case must argue that the adaptation is "reasonably" (to the extent practically possible) safe based on evidence available at design-time {TIII.B2}. For example, before deployment, extensive simulation studies can be used to determine how the system might adapt to various operating conditions. Third, the safety case must show that the programmatic procedures used by the managing system at run-time to determine whether a given adaptation option is safe are suitable, even if they have limitations{TIII.B3}. Fourth, the safety case should show that the managing system will not apply an adaptation option that it determines to be unsafe {TIII.B4}. Finally, the safety case must show that the managing system detects and responds to changes in SPIs monitored by the managing system {TIII.B5}.

The dynamic portion of the safety case must account for two items. First, at run-time the safety case must be adapted to include the evidence generated by the managing system when assessing the safety of the chosen adaptation option{TIII.B6}. Second, the managing system should monitor SPIs and update the safety case (in real-time) with this data {TIII.B7}.

*1) Type III Example:* To illustrate Type III adaptation, consider an instance of the water heater that, instead of using a PID controller, uses a neural network to determine the level of heating. The neural network consumes the current and historical values of system state parameters, including the inflow temperature, inflow rate, outflow temperature, external temperature. The managing system adapts the neural network by changing both the weights of the network and the hyperparameters (number of layers, layer size, activation fucntions, etc.). Since the set of possible neural networks is very large, it is not practical to enumerate all of the options or show that each one is safe at design-time. Instead, at run-time the managing system applies a programmatic decision procedure to determine whether a candidate neural network is safe. This procedure consists of a suite of simulation scenarios that probe for different (possible hazardous) behaviours. If the neural network successfully completes the simulations, then the managing system applies the adaptation by: 1) updating the neural network in the controller; 2) updating the safety case, within the knowledge repository of the managing system, to include the latest simulation results; and 3) resetting the SPI monitoring.

One of the monitored SPIs is the duration of time where outflow water temperature was within 5% of the hazardous limit (90 °C). If this duration exceeds a pre-determined threshold (e.g., 60 seconds over the last hour) then it suggests that the system is coming close to the hazardous limit. If this occurs, the managing system performs a fail-safe action and resets the controller's configuration.

V. DISCUSSION

Thus far, this paper has provided definitions for *safety-critical self-adaptive systems* and *safe adaptation*. These were used to formulate a taxonomy for classifying adaptations based on their impact on the safety of the managed system and the system's safety case. This section provides additional context for the taxonomy by discussing its applicability to an industrial setting, the impact of learning in the classification of adaptations, and threats to validity.

*A. Industrial Applicability*

The taxonomy for classifying adaptations may be applied in at least two ways in an industrial context: i) to guide designers developing an assurance case and ii) to guide regulators assessing self-adaptive systems.

First, for each type of adaptation, the taxonomy provides "obligations" that could be used by designers to guide the development of their safety case(s). In practice, concrete guidelines are adopted to reduce the burden on designers to "re-invent the wheel" for each new project and allow them instead to focus on the important project specific details. Assurance case templates (also referred to as "patterns") are a common method capturing recurring assurance arguments for re-use or for communicating a novel argumentation approach [12]. For example, Koopman and Osyk described an assurance case template for public road testing of autonomous vehicles which can be used by organizations when preparing their own road testing programs [23]. The taxonomy and its obligations may be used in a similar manner.

Second, a common regulatory strategy for safety-critical systems is to assign "classes" or "levels" to systems based on their assessed level of risk. Higher levels of risk demand engineering activities with increasing rigour be applied to the system.

*Table 1 – Summary of taxonomy (criteria and obligations for classifying adaptations in safety-critical self-adaptive systems.*

| Type | System Criteria (for each adaptation) | Safety Case Obligations (for each adaptation) |
|---|---|---|
| **Type 0** *Non-Interference* | T0.C1 – Adaptation does not affect the safety-critical functions of the managed system. | T0.B1 – Argue that the adaptation does not interfere (or otherwise affect) with the safety-critical functions. <br> T0.B2 – The safety case is defined at design-time and does not change at run-time, i.e., the safety case static. |
| **Type I** *Static Assurance* | TI.C1 – Adaptation affects safety-critical functions of the managed system. <br> TI.C2 – All adaptation options are defined at design-time. <br> TI.C3 – All adaptation options are shown to be safe at design-time. | TI.B1 – Argue that the managing system only executes adaptations defined at design-time. <br> TI.B2 – Argue that each adaptation option is safe within the entirety of the system's operational domain. <br> TI.B3 – Statically argue that the managing system safely executes the adaptation action. <br> TI.B4 – The safety case is defined at design-time and does not change at run-time, i.e., the safety case static. |
| **Type II** *Constrained Assurance* | TII.C1 – Adaptation affects safety-critical functions of the managed system. <br> TII.C2 – All adaptation options are defined at design-time. <br> TII.C3 – All adaptation options are shown to be safe at design-time subject to constraints on the operational domain. <br> TII.C4 – Adaptation imposes constraints on the operational domain and the system may only be operated within a restricted domain defined by the adaptation option; the constraints are reflected in the safety case. <br> TII.C5 – Adaptations monotonically increase constraints on the operational domain. | TII.B1 – Statically argue that the managing system only executes adaptations defined at design-time. <br> TII.B2 – Statically argue that the defined adaptation options are safe subject to assumptions on the operational domain. <br> TII.B3 – Statically argue that the managing system safely executes the adaptation action. <br> TII.B4 – Dynamically argue that, if an adaptation option is applied, then the current operational domain satisfies the constraints associated with that option. <br> TII.B5 – Dynamically argue that violations of the current operational constraints are either safely handled or not possible. |
| **Type III** *Dynamic Assurance* | TIII.C1 – Adaptation affects safety-critical functions of the managed system. <br> TIII.C2 – The set of adaptation options is not defined at design-time and so cannot be shown to be safe at design-time. <br> TIII.C3 – Adaptation options are assessed at run-time by the managing system and the results are incorporated into a dynamic safety case as part of the adaptation operation. <br> TIII.C4 – The safety case is stored as part of the knowledge repository within the managing system. <br> TIII.C5 – Uncertainty in safety arises from the limitations of the assessment methods used by the managing system to determine if an adaptation option is safe. | TIII.B1 – Statically argue that the managing system safely executes the adaptation action. <br> TIII.B2 – Statically argue that the adaptation is "reasonably" safe based on evidence available at design-time. <br> TIII.B3 – Statically argue that the procedure(s) used by the managing system at run-time to determine whether a given adaptation option is safe are appropriate. <br> TIII.B4 – Statically argue that the managing system will not apply an adaptation option that it determines to be unsafe. <br> TIII.B5 – Statically argue that the managing system detects and responds to changes in monitored safety performance indicators. <br> TIII.B6 – Dynamically argue, based on evidence generated by the managed system's assessment, that the selected adaptation option is safe. <br> TIII.B7 – Dynamically argue, based on real-time data from safety performance indicators, that system's operation continues to be safe. |

For example, ISO 26262 uses Automotive Safety Integrity Levels (ASILs) A (lowest risk) through D (highest risk) [7]. A similar strategy could be employed by regulators assessing safety-critical self-adaptive systems. The types described by the taxonomy have increasing levels of uncertainty associated with safety with Type 0 having the least uncertainty and Type III have the most. Regulators might apply the taxonomy to determine the level of uncertainty arising from an adaptation. This would not replace existing level-of-rigour approaches for safety-critical systems. Instead, it could be added as "another dimension" of existing risk assessment procedures.

*B. Learning*

Machine learning and artificial intelligence ("learning") techniques may be applied to self-adaptive systems [1, 24]. Learning may be used in the managed or managing systems. In the managed system, the primary function of the system may be realized (or supported) by learning. In the managing system, learning is often used to: update and improve models in the knowledge repository, update adaptation rules or policies, or improve the selection of adaptation options [24].

Despite the importance of learning techniques to autonomous and adaptive systems, it is interesting to observe that the taxonomy in Section IV does not mention learning as a criterion for determining the type of an adaptation. Whether the managing or managed systems make use of learning has no bearing on the classification of an adaptation. In some sense this is surprising, since assuring learning-based systems are typically thought of as more difficult than assuring conventional software, which suggests that the type classifications for a self-adaptive system using learning techniques should tend towards Type III, where there is less certainty about safety. However, it is not whether learning is used, but the role of learning within the system that matters.

Consider the instance of the water heater described in Section IV.D. This instance uses a reinforcement learning technique to *generate* new adaptation options (neural network weights for the water heater's controller) at run-time. The fact that learning is used contributes to uncertainty about whether the generated adaptation options are actually safe, which leads to a Type III classification. However, learning can also be used in a manner that does not affect safety. Suppose instead that a pre-defined set of adaptation options for a PID controller are used, like in the water heater instance in Section IV.B: each of these options is shown in advance to be safe. Learning could still be used to select the best adaptation option for a given circumstance. In this case, the learning cannot affect the safety of the system, though it might affect performance or reliability if a sub-optimal adaptation option is chosen.

In summary, learning should be viewed as an "orthogonal" concept to the safety classification of adaptations. While the use of learning can certainly have an impact on the classification of an adaptation, it is not in itself a criterion for classification.

*C. Threats to Validity*

Since this is a concept paper, the focus has been to formulate the definitions related to safety-critical self-adaptive systems and to describe a taxonomy for classifying adaptations. To date, validation of the taxonomy is limited to the illustration of the adaptation types using the water heater example. This mode of validation is clearly insufficient to show that the taxonomy can be generalized to a wide range of safety-critical self-adaptive systems. There are at least four ways in which the proposed taxonomy could be insufficient. First, there might be other types of adaptations that are not accounted for in the taxonomy. Second, even if the taxonomy covers all adaptation types in concept, the criteria might not be expressed in a manner (too precise or too abstract) that permits classification of adaptations in practice. Third, the obligations might be missing an important idea or are under prescriptive such that safety cases incorporating them are not convincing to stakeholders. Finally, the obligations might be overly prescriptive or overly onerous such that they cannot reasonably be used in practice. Further validation of the taxonomy is necessary.

From a safety perspective, the water heater example used to illustrate the types of adaptation in IV is relatively simple. This example was selected so that concepts could be discussed without getting lost in technical details of a more sophisticated system. Indeed, the simplest option to mitigate the over-temperature hazard would likely be to use an independent safety monitor as shown in Figure 3. While the safety monitor pattern is a "tried and tested" design for safety-critical systems (and we would likely advocate for its use if this was a real-world safety-critical water heating system), simply employing a safety monitor would deprive us of an opportunity to explore the more interesting Type I, II, and III adaptations. The water heater example should be regarded as a proxy for more sophisticated systems where an independent safety monitor is not feasible or where the safety monitor is itself subject to adaptation, such as the autonomous parcel delivery robot.

## VI. RELATED WORK

Assurance of self-adaptive systems, including safety-critical systems, has been addressed by many authors. The existing approaches are generic and may be used to assure a range of properties such as reliability, availability, performance, and of course safety. When safety is considered, it is often used as a motivation for assuring a system. However, to our knowledge, there is no work that addresses safety at a conceptual level by providing definitions, frameworks, or taxonomies for safety. Some recent works have begun to address safety in more abstract terms [25, 26], but have not yet offered concrete definitions of safety for self-adaptive systems. Instead of reviewing related definitions or taxonomies, this section reviews existing approaches to assuring self-adaptive systems that make use of assurance cases, which are a foundational aspect of the taxonomy.

Calinescu *et al.* developed a method called ENTRUST (ENgineering TRUstworthy Self-adaptive sofTware) for creating dynamic assurance cases for self-adaptive systems [6]. The ENTRUST methodology uses a combination of design-time and run-time modelling and verification techniques to assure a self-adaptive system. At design-time a partially instantiated safety case is created. Then, at run-time, the managing system generates evidence using the modelling and verification techniques and updates the remaining (dynamic) elements of the assurance case. ENTRUST was demonstrated on two case studies: an autonomous underwater vehicle and a stock trading platform. Though the ENTRUST method could be applied to a safety-critical system the method is generic and could be applied to assure any property of a self-adaptive system. The case studies described by Calinescu *et al.* appear to focus mainly on reliability and performance, rather than safety.

Cheng *et al.* describe a framework called AC-ROS (Assurance Cases for Robot Operating System) that uses an assurance case, developed at design time using GSN, to automatically reason about whether a robot is satisfying its safety objectives [27]. In AC-ROS, the solution nodes (leaves) of the GSN tree contain "utility functions" that the Monitor phase of the MAPE uses to monitor properties of the environment. The Analyze phase uses the structure of the GSN tree to determine whether the safety case is currently valid. If the case is not valid then an adaptation option is selected from a pre-defined list (e.g., switching operating modes of the robot) and executed. Cheng *et al.* demonstrate AC-ROS using EvoRally, a small four-wheeled vehicle intended for autonomous systems research that runs the ROS.

Mirzaei *et al.* consider the use of dynamic safety cases to assure self-adaptive systems of systems [28]. Like Calinescu *et al.* and Asaadi *et al.*, Mirzaei *et al.* identify the importance of a dynamic safety case that contains both static and dynamic aspects with the latter being generated at run-time. They suggest that dynamic safety cases be combined with modularity practices to manage the complexity that arises in self-adaptive systems of systems and call these Dynamic Modular Safety Cases (DMSCs).

Jahan *et al.* focus on self-adaptative systems that use security assurance case as a run-time model to determine whether security objectives are satisfied [29, 30]. They have defined an assurance case template for arguing that security controls (e.g., NIST 800-53 controls) are correctly implemented and define adaptation operators that a managing system can use to change a dynamic assurance case at run-time [29] . Then, in [30] Jahan *et al.* apply dynamic change impact assessment to an assurance case maintained by the managing system. They compose multiple security assurance cases, each arguing that a specific security control is realized by the system, into a security control network (SCN) model. The impact of an adaptation on the SCN is assessed and used by the managing system to select the least impactful adaptation. Impacts are assessed using structural graph metrics, such as degree of nodes affected in the assurance case. They demonstrate their approach on smart inventory management system.

Asaadi *et al.* describe an architecture for "trustworthy autonomy" based on the notion of a dynamic assurance case [20]. Similarly to Calinescu *et al.*, they describe an assurance case as consisting of both static and dynamic elements; however, they differ on the role of the dynamic aspects. Asaadi *et al.* propose that the dynamic elements of the case capture "assurance measures" that capture the degree of confidence in the case at run-time. They propose a reference architecture that uses the dynamic assurance case to determine whether to perform "contingency management" operations on the managed system. Asaadi *et al.* demonstrate their concept on an autonomous taxiing application for aircraft. Though their work is framed using the language of autonomous systems (not self-adaptive systems) their proposed reference architecture is very similar to that of a self-adaptive system and so it is included here. This work extends the concept of dynamic assurance proposed by Denney *et al.* in [19].

## VII. Conclusion and Next Steps

To date, researchers have focused on assuring properties of self-adaptive systems like reliability, performance, availability, and safety. Though many authors have written about safety-critical self-adaptive systems, no concrete definitions for this class of system have been provided. To address this gap, this concept paper gives definitions for the terms *safety-critical self-adaptive system* and *safe adaptation*. These definitions were used to formulate a taxonomy that classifies adaptations based on the impact they have on the system's safety. The next step for this line of inquiry is to validate the proposed taxonomy to demonstrate that it is capable of classifying all types of safety-critical self-adaptive systems and that the obligations imposed by the taxonomy are appropriate using a combination of systematic literature reviews and case studies.